\documentstyle[11pt,IAUS212,twoside,epsf]{article}

\def\edcomment#1{\iffalse\marginpar{\raggedright\sl#1\/}\else\relax\fi}
\reversemarginpar

\begin{document}
\vspace*{1cm}
\title{Stellar Parameters of Wolf-Rayet Stars from Far-UV to Mid-IR Observations}
 \author{Paul A. Crowther}
\affil{Dept of Physics \& Astronomy, University College London, Gower St., 
London WC1E 6BT, UK, pac@star.ucl.ac.uk}

\begin{abstract}
Recent results for Galactic and Magellanic Cloud Wolf-Rayet stars are
summarised based on line blanketed, clumped model atmospheres together with
UV, optical and IR spectroscopy. The trend towards earlier WN and WC spectral
types with decreasing metallicity is explained via the sensitivity of 
classification diagnostics to abundance/wind density, such that WR mass-loss
rates are metallicity dependent. Pre-supernovae masses for WC stars are 
determined, in reasonable agreement with CO-cores of recent Type-Ic SN.
\end{abstract}

\section{Recent Observational and Theoretical Progress}


This article will focus on recent determinations of physical parameters 
for Galactic and Magellanic Cloud WR stars from UV to IR diagnostics.
Observationally, the {\it Far-Ultraviolet Spectroscopic Explorer 
(FUSE)} has provided an impressive database of $\lambda$912--1187\AA\ 
spectroscopy for Galactic and Magellanic Cloud WR stars (see Willis et al. 
these proc.) to  supplement previous UV datasets obtained with the
{\it International Ultraviolet Explorer (IUE)} and 
{\it Hubble Space Telescope (HST)}. At longer wavelengths, {\it Infrared 
Space Observatory (ISO)} observations of WR stars have now been analysed.
Much of the recent observational progress with Wolf-Rayet stars 
has involved the acquisition of high quality optical spectroscopy
for individual stars beyond the Magellanic Clouds (Drissen, these proc.), 
plus X-ray spectroscopy of single and binary
WR stars (e.g. Skinner et al. 2002) neither of which topics will
be discussed here.


Theoretical developments in the last few years have been more steady,
with the (laborious) implementation 
of line blanketing into codes by elements other than CNO and Fe,
which had already been discussed at the last hot star 
beach symposium, IAU Symp. 193. The major change has been the
widespread use of such codes to analyse individual stars within
a range of galaxies. At present, there are a variety of model atmosphere
codes which consider sphericity and line blanketing and fall into two
main types, outlined below.

CMFGEN (Hillier \& Miller 1998) and the Gr\"afener et al.
(2002) code make use of variants of the super-level approach to 
incorporate the effect of tens of thousands metal lines on the atmospheric
structure within the radiative transfer code. 
CMFGEN can now simultaneously consider blanketing by individual
ions of up to 30 elements, including C, N, O, Ne, Si, S, Ar, Ca, Fe and Ni
(Hillier, these proc.), whilst Gr\"afener et al. consider CNO, Si plus
Fe-group elements (Sc to Ni) grouped together in a single generic atom.
This approach suffers the least number of approximations, but 
remains computationally demanding. Recent test calculations for 
early-type WC stars show (perhaps surprisingly!) good consistency between 
these two codes, including ionizing fluxes.

Alternatively, Schmutz (1997) and ISA-wind (de Koter et al. 1993, 1997) 
use separate codes to solve the radiative transfer problem and 
line blanketing, the latter making use of Monte Carlo techniques. 
The method had the great computational 
advantage that complete intensity-weighed effective opacity factors can be
calculated separately from the transfer problem. 
On the negative side, the ionization and excitation equilibrium of metal 
species is approximate, dictating which lines are efficient at 
capturing photons for each point in the atmosphere.
Test calculations for a late-type WN star
between CMFGEN and ISA-wind show excellent agreement in derived
stellar parameters, but rather
poorer agreement for ionizing fluxes (Crowther et al. 1999).
\begin{table}
\caption{\small Recent revisions in the derived 
stellar parameters (clumped in bold with $f$=0.1) for HD~165763 (WR111, WC5)
due to the incorporation of metal line blanketing.\label{table1}}
\begin{center}
\begin{tabular}{lcllcl}
 \hline
$T_{\ast}$ & $\log L$ & $\log \dot{M}$ &Elements & Blank?& Reference \\
kK         &  $L_{\odot}$   & $M_{\odot}$yr$^{-1}$ &included& &\\
\hline
35         & 4.6                  & $-4.6$ & He & no& Schmutz et al. 1989\\
59         & 5.0                  & $-4.4$ & He, C & no & Hillier 1989\\
90  &  5.3            & {\it\bf $-$4.8} & He, C, O, Fe & yes&
Hillier \& Miller 1999 \\
85   & 5.45           & {\it\bf $-$4.9} & He, C, O, Fe-group & yes
& Gr\"{a}fener et al. 2002\\   
\hline
\end{tabular}
\end{center}
\end{table}

The main effect of blanketing is to re-distribute extreme UV flux to 
longer wavelengths, reducing the ionization balance in the atmosphere,
such that higher stellar temperatures (and luminosities) are required
to match observed line profile diagnostics relative to unblanketed 
studies. This is illustrated in Table~1 for the prototypical 
Galactic early-type WC star HD~165763 (WR111) whose derived
stellar luminosity has increased by a factor of 5--7 over the past decade. 
Differences in luminosities for HD~165763 between the recent studies of 
Hillier \& Miller (1999) and Gr\"afener et al. (2002) most likely result
from the inclusion of additional blanketing elements, which CMFGEN 
now routinely handles. Recent revisions to temperatures and luminosities of
O supergiants have acted in the opposite sense, relative to
previous plane-parallel unblanketed model analyses, such that common 
techniques are now employed throughout for O and WR stars
(e.g. Crowther et al. 2002ab).

Clumping is now routinely, albeit approximately, handled in WR 
model atmospheric studies via radial dependent volume filling factors, $f$. 
Constraints on $f$ can be obtained from comparisons between red electron
scattering wings and observations (e.g. Hillier 1991), although exact 
determinations generally prove elusive due to line blending, particularly in 
WC stars. 
Generally, $f\sim$0.05--0.25
provide reasonable matches to observed line profiles (e.g. Hamann \& Koesterke
1998), such that global mass-loss
rates are reduced by a factor of $1/\sqrt{f} \sim 2-4$ relative to smooth models.
The majority of line profiles behave rather insensitively provided $\dot{M}/\sqrt{f}$
remains constant with some exceptions (e.g. Herald et al. 2001).
In WC stars, the line centre of UV resonance lines of C\,{\sc iii-iv} reacts to
changes in $f$, 
providing additional constraints on the filling factor (Crowther et al. 2002a).



\section{WN properties}

Although large samples of WN stars have not yet been thoroughly analysed 
using recent line blanketed codes, stellar temperatures of WN stars 
range from 30kK (at WN10), to $\sim$40kK (at WN8) and approaching 100kK 
for early-type WN (WNE) stars. 
Increased stellar temperatures, particularly for WNE stars, 
with correspondingly smaller radii,
brings atmospheric models into much closer agreement with direct 
determinations from short period WN+O binaries, such as V444 Cyg 
(e.g. Moffat \& Marchenko 1996). 
Clumped mass-loss rates from optical diagnostics generally lie
in the range $\dot{M}=10^{-5.5 \ldots -4.5} M_{\odot}$ yr$^{-1}$.
Mass-loss rates can also be obtained from radio determinations,
although these too are subject to uncertainties in volume filling
factors, and are limited to stars within a few kpc. 
Typical wind velocities span a wide range, well correlated with spectral type:
300\,km\,s$^{-1}$ at WN10 to $\sim$2000\,km\,s$^{-1}$ at WN3-4.

Relative to recent unblanketed results,
Herald et al. (2001) obtained a smaller
core radius and slightly enhanced luminosity 
for HD~96548 (WR40, WN8) allowing for blanketing including Ne, Ar, Ca and 
Fe,  such that the heavily blended  Fe\,{\sc iv-v} forest in the UV is 
very well reproduced. 
This naturally leads to the potential of late-type WN (WNL) stars as providing
diagnostics of heavy elemental abundances from UV and far-UV ({\it FUSE}) spectroscopy. 
Morris et al. (2000) carried out a 
combined optical--infrared analysis of another WN8 star AS~431 (WR147)
revealing overall excellent consistency between visual, near-IR and mid-IR
{\it ISO} line diagnostics, such that reliable stellar parameters may be
obtained from solely IR observations, providing all necessary diagnostics
are covered. Meanwhile, Crowther et al. (1999) used the H\,{\sc ii} ejecta nebula 
M1--67 as a sensitive diagnostic of the ionizing flux distributions predicted
by analyses of the central star 209~BAC (WR124, WN8) using
CMFGEN and ISA-wind, favouring the former in this case.



In general, WNL stars are relatively H-rich (H/He$\sim$0.5-2 by number), whilst
WNE stars do not contain hydrogen, although exceptions do occur. The question of
whether weak-lined WN5--6 stars in R136 are moderately enriched in helium (H/He$\sim$3,
Crowther \& Dessart 1998) or unprocessed (H/He$\sim$10, de Koter et al. 1997) remains
open.  Where studies of WN stars have been carried out, N is extremely enriched and
C depleted, consistent with CN-cycle processed material, with O generally more difficult
to constrain (e.g. Herald et al. 2001). In principal, other elemental abundances 
may be obtained from UV synthesis, but the most direct method is via analysis of
mid-IR fine structure lines. Morris et al. (2000) obtained lower limits of $\sim$0.5 solar
for the abundance of Ne, S and Ca in AS~431 from ISO observations, with individual
determinations sensitive to the adoption of smooth or clumped winds (see e.g. 
J.D.~Smith, these proc.).

WN spectral types mimic underlying stellar temperatures rather well, at least 
within a particular (metallicity) environment. However, since WN subtypes depend 
on the relative strength of nitrogen emission lines,  samples within 
different galaxies do not necessarily behave in the same manner. 
This is because N\,{\sc iii-v} classification lines have differing 
sensitivities to abundance which will be a factor of five times lower for a
SMC WN star than its counterpart in the Solar neighbourhood. Crowther 
(2000) demonstrated that a WN5 star in the SMC would have
a later spectral type in the Milky Way if its 
stellar parameters were kept {\it constant}, and an earlier spectral
type at lower metallicity.
This tendency broadly explains the observed  trend 
earlier WN spectral types when one compares the statistics of WN stars 
in the Galaxy and Magellanic Clouds.

\section{WC properties}

Due to the overwhelming effect of metal line blanketing on the atmospheric
structure of WC stars, considerable effort has recently gone into their analysis.
Stellar temperatures range from $\sim$45kK (WC9) to $\sim$60kK (WC8) to $\sim$100kK (WCE)
and likely well in excess of this value for some WO stars. Mid-IR {\it 
ISO} observations
(e.g. De Marco et al. 2000) and far-UV {\it FUSE} spectroscopy (e.g. 
Crowther et al. 2000) have
been combined with the usual UV and optical diagnostics 
to derive stellar parameters. The use of similar techniques
for Galactic (e.g. Dessart et al. 2000) and 
Magellanic Cloud (Crowther et al. 2002a)
WC stars permits their relative properties to be investigated in detail. 
Clumped WC mass-loss  rates lie in the range $\dot{M}=10^{-5.0 \ldots -4.4} 
M_{\odot}$  yr$^{-1}$. As for WN stars, wind velocities are well 
correlated with spectral type:
1100\,km\,s$^{-1}$ at WC9, to 3000\,km\,s$^{-1}$ at WC4.

Carbon abundance determinations of WC stars have long used the He\,{\sc ii} $\lambda$5411
and C\,{\sc iv} $\lambda$5471 recombination lines. However, conflicting results may result
if limited model atoms are used, or if trace elements are neglected, such that Gr\"afener
et al. (1998) obtained C/He=0.32 by number (40\% by mass) for HD~32125 (WC4), 
in contrast with the recent determination of C/He=0.13 by number
(25\% by mass) from Crowther et al. (2002a). Carbon abundances in Galactic and LMC WC stars
span a similar range, 0.1 $\leq$ C/He $\leq$ 0.4, such that spectral types are 
{\it not} determined by carbon abundances, in conflict with predictions by 
Smith \& Maeder (1991).
Hamann et al. (these proc.) suggest a relatively uniform carbon abundance
of C/He$\sim$0.25 for most Galactic WCE stars. Sand~2 (LMC, WO) does appear to have a
higher C/He ratio than WC stars, but determinations for such stars are hindered by severe
blending caused by it's broad lines.

Oxygen abundances are rather more difficult to tightly constrain, since most diagnostics
lie around $\lambda$3000\AA, and span a much wider range of ions -- O\,{\sc iii-vi} in most
WC stars -- than carbon. Indeed, the classification line at $\lambda$5590 is formed from 
a blend of O\,{\sc iii} and O\,{\sc v} lines. Consequently, O/He determinations can 
generally not be obtained better than a factor of $\sim$2. Oxygen in Galactic and LMC
WC stars ranges from 0.02 $\leq$ O/He $\leq$0.1, again with the probable exception of 
Sand~2 (Dessart et al. 2000; Crowther et al. 2000, 2002a). As for WN stars, other 
elemental abundances are most readily determined from mid-IR fine structure lines, provided
clumping factors can be determined. For $\gamma$ Vel (WC8+O, WR11), 
Dessart et al. (2000) find that S is within 20\% of the  
cosmic value, whilst Ne is enhanced by a factor of $\sim$8, in reasonable 
agreement with recent evolutionary predictions. 



If WC spectral types do not result from chemical abundance variations, 
why are LMC WC stars systematically of earlier spectral type than Galactic WC stars? 
Temperature certainly plays a role, but for an answer we must first consider the WC
classification lines -- C\,{\sc iii} $\lambda$5696 and C\,{\sc iv} 
$\lambda\lambda$5801-12 -- in greater detail. The upper level of $\lambda$5696 has an
alternative decay via $\lambda$574 with a branching ratio of 147:1. Consequently,
$\lambda$5696 only becomes strong when $\lambda$574 is optically thick, i.e. if
the stellar temperature is low {\it or} the wind density is high. A comparison
between Galactic WC5--7 stars and LMC WC4 stars reveals that their temperatures are
comparable, such that the wind densities of the Galactic sample must be higher than
the LMC stars. Crowther et al. (2002a) demonstrate that this is indeed the case, such 
that the principal difference amongst WCE stars is differing wind densities -- high for
WC6-7 stars and low for WC4 stars, with 
\begin{equation}
\label{mdot}
\log (\dot{M}) = 1.38 \log (L/L_{\odot}) - 12.35
\end{equation}
for the LMC stars, and Galactic WC stars $\sim$0.25 dex higher, in good agreement
with the {\it generic} WR mass-loss luminosity calibration of Nugis \& Lamers (2000).


\begin{figure}[thbp]
\epsfxsize=8.0cm \epsfbox[-85 210 415 610]{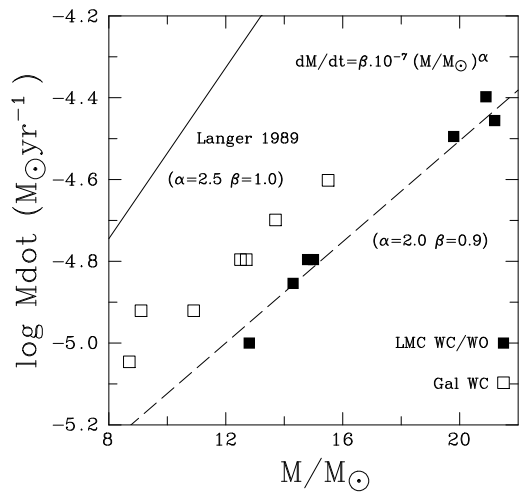}
\caption{\small Comparison between WC masses and mass-loss rates
for Galactic (open) and LMC (filled) stars derived by recent
atmospheric calculations (see Crowther et al. 2002a) and the
calibration of Langer (1989) for different $\alpha$ and $\beta$ (see text)}
\label{mdot}
\end{figure}

\section{A Metallicity dependence for Wolf-Rayet stars?}

A wind density origin for the subtype distribution amongst WC stars in Local Group
galaxies can be understood if the mass-loss rates of Wolf-Rayet stars are metallicity
dependent. The difference between Galactic and LMC WCE stars amounts to 
only a factor of two, which is consistent with a metallicity dependence of 
$Z^{0.5-0.7}$ as predicted
by radiation driven winds for O stars (e.g. Vink et al. 2001). Reduced
wind densities for WC stars
 at lower metallicity, preferentially affecting C\,{\sc iii} $\lambda$5696, 
naturally explains the trend towards WC4 (and WO) stars, 
as is observed in the Magellanic Clouds and IC10 (Crowther et al. these
proc). Although the atmospheres of WC stars
 are composed of mostly He, C and O, the heavier elements 
(Ne, Ar, Fe, Ni...) initiate their winds, as in O stars. 

A WR metallicity dependence is not currently adopted in
evolutionary calculations, which generally typically adopt $\dot{M}=\beta
\times10^{-7} (M/M_{\odot})^\alpha$ for all H-free WR  stars (Langer 1989). 
For WC stars Langer adopted $\alpha$=2.5 and $\beta$=1.0. This relationship is
compared to recent mass-loss rates and masses for Galactic and LMC WC 
stars in Fig.~1, revealing very poor agreement. 
LMC WC stars are well matched using $\alpha$=2.0 and $\beta$=0.9, similar
to that discussed by Cherepashchuk (2001) from independent methods.

If WC stars are metallicity dependent, one would expect the same of WN stars. However,
results to date for WN stars has been less clear (Crowther \& Smith 1997; 
Hamann \& Koesterke 2000), most likely due to greater scatter in wind properties at
a given spectral type. Nevertheless, the trend is also towards 
weaker winds, as is seen most readily via the single WN stars in the SMC.
The question of whether WR mass-loss rates are metallicity dependent has
a major impact on the hardness of their 
ionizing flux distributions, as discussed by L.J.~Smith et al. (these
proc.). Denser winds cause softer Lyman continuum fluxes, which would
imply that few WR stars produce nebular He\,{\sc ii} ionizing photons at high
metallicity, with the opposite true at low metallicities.


\section{Pre-Supernova WR Masses}

Recent theoretical progress in our analysis of Wolf-Rayet stars has led us
to the claim that derived stellar parameters are `robust'. Since there is
a well established theoretical mass-luminosity relation for hydrogen-free
WR stars, those which are members of close period binaries offer us
the possibility to check derived spectroscopic luminosities. De Marco et
al. (2000) obtained a stellar luminosity of the WC8 component of $\gamma$ Vel
which implies a mass of 9$M_{\odot}$ (via the Heger \&  Langer 1996 relationship), 
in remarkable
agreement with 9.5$M_{\odot}$ derived from the binary orbit. 

We can employ evolutionary models to obtain remaining lifetimes, which
together with an appropriate mass-loss luminosity relation (e.g. Eqn.~1),
permits an estimate of the final, pre-SN mass. This approach can be 
followed for all WR stars. However, uncertainties in evolutionary 
models and lack of recent results for WN stars suggest that WC stars 
represent our best 
candidates for the determination of pre-SN masses. Taking the 
LMC WC4 star HD\,32402 
as an example, Crowther et al. (2002a) derive a current mass of 21$M_{\odot}$,
a remaining lifetime in the range 1--4$\times10^{5}$yr, and so a pre-SN mass of
14$M_{\odot}$ (non-rotating model) or 19$M_{\odot}$ (rotating model). From a
sample of Galactic WC stars at known distance, final pre-SN masses lie in 
the range 7--14$M_{\odot}$, versus 11--19$M_{\odot}$ using LMC stars.

A number of Type-Ic supernovae that have been studied in the past few years
have probable WR precursors, including SN 1998bw, for which a CO (ejected) core
mass of 13.8$M_{\odot}$ was derived by Iwamoto et al. (1998). This agrees rather
well with the expected pre-SN masses of LMC WC stars. There has been considerable
interest in SN 1998bw since it was also GRB\,980425. 
SN 2002ap is another nearby Type-Ic for which a WR star is one of the few likely 
progenitors (Smartt et al. these proc.). Unfortunately, most hypernovae or `collapsar'
models require rapid rotation, whilst rotation rates are lacking for almost all WR
stars. The general consensus is that most WR stars have spun-down -- indeed practically
all WC stars are thought to be spherical, as inferred from spectropolarimetry of their winds
(Harries et al. 1998). 

\acknowledgements

Many thanks to the Royal Society for continued financial support, and
to my collaborators without whom the work presented here
would not have  been possible, John Hillier and Luc Dessart in particular.


\vspace*{-0.1cm}
\begin{references}
{\small
\reference Cherepashchuk A.M., 2001, Astron. Rep. 45, 120
\reference Crowther P.A., 2000, A\&A 356, 191
\reference Crowther P.A., Dessart L. 1998, MNRAS 296, 622
\reference Crowther P.A., Smith L.J., 1997, A\&A 320, 500
\reference Crowther P.A., Pasquali A., De Marco O., et al. 1999, A\&A 350, 1007
\reference Crowther P.A., Fullerton A.W., Hillier D.J. et al. 2000, ApJ 538, L51
\reference Crowther P.A,, Dessart, L., Hillier D.J. et al., 2002a,
A\&A in press (astro-ph/0206233)
\reference Crowther P.A., Hillier D.J., Evans C.J. et al., 2002b, ApJ in press 
(astro-ph/0206257)
\reference de Koter, A., Schmutz W., Lamers H.J.G.L.M., 1993, A\&A 277, 561
\reference de Koter, A., Heap, S.R., Hubeny, I.  1997, ApJ 477, 792
\reference De~Marco, O., Crowther P.A., Schmutz W., et~al.,   2000, A\&A 358, 187
\reference Dessart L., Crowther P.A., Hillier, D.J. et al., 2000, MNRAS 315, 407
\reference Gr\"{a}fener, G., Hamann, W-.R., Hillier, D.J., Koesterke, L., 1998, A\&A, 329, 190
\reference Gr\"{a}fener, G., Koesterke, L., Hamann, W-R., 2002, A\&A 387, 244
\reference Hamann W.-R., Koesterke L., 1998, A\&A 335, 1003
\reference Hamann W.-R., Koesterke L., 2000, A\&A 360, 647
\reference Harries T.J., Hillier D.J., Howarth I.D., 1998 MNRAS 296, 1072
\reference Heger A., Langer N., 1996 A\&A 318, 421
\reference Herald J.E., Hillier D.J., Schulte-Ladbeck, R.E., 2001, ApJ 548, 932
\reference Hillier D.J.,  1989, ApJ 347, 392
\reference Hillier D.J., 1991, A\&A 247, 455
\reference Hillier D.J., Miller D.L., 1998, ApJ, 496, 407
\reference Hillier D.J., Miller D.L., 1999, ApJ, 519, 354
\reference Iwamoto K., Mazzali, P.A., Nomoto K., et al. 1998, Nat 395, 672
\reference Langer N., 1989, A\&A 220, 135
\reference Moffat A.F.J., Marchenko S.V., 1996, A\&A 305, L29
\reference Morris P.W., van der Hucht K.A., Crowther P.A. et al. 2000, A\&A 353, 624
\reference Nugis T., Lamers H.J.G.L.M. 2000, A\&A 360, 227
\reference Schmutz W.,  1997, A\&A 321, 268
\reference Schmutz W., Hamann, W.-R., Wessolowski, U. 1989, A\&A 210, 236
\reference Skinner S.L., Zhekov, S.A., G\"udel, M., Schmutz W., 2002, ApJ 572, 477
\reference Smith L.F., Maeder A.  1991, A\&A 241, 77
\reference Vink J.S., de Koter A., Lamers H.J.G.L.M, 2001, A\&A 369, 574
}
\end{references}
\end{document}